\documentclass[aip,apl,twocolumn,reprint]{revtex4-1}
\usepackage{graphicx}
\usepackage{dcolumn}
\usepackage{bm}
\usepackage{array}
\usepackage[toc,page]{appendix} 

\usepackage[utf8]{inputenc}
\usepackage[T1]{fontenc}
\usepackage{color,soul}
\usepackage{tabularx}
\usepackage{amsmath}
\usepackage[normalem]{ulem}

\begin{document}
\title{Highly Tunable Photon-Magnon Coupling Governed by Collective-Mode Profiles in Reconfigurable Dielectric-Resonator Arrays}

\author{Shihao Zhou}
\affiliation{Department of Physics and Astronomy, University of North Carolina at Chapel Hill, Chapel Hill, NC 27599, USA}
\author{Edward Kwao}
\affiliation{Department of Electrical and Computer Engineering, North Carolina A$\&$T State University, Greensboro, NC 27411, USA}
\author{Junming Wu}
\affiliation{Department of Physics and Astronomy, University of North Carolina at Chapel Hill, Chapel Hill, NC 27599, USA}
\author{Abhijit Shinde}
\affiliation{Department of Electrical and Computer Engineering, North Carolina A$\&$T State University, Greensboro, NC 27411, USA}
\author{Yin Lyu}
\affiliation{Department of Physics and Astronomy, University of North Carolina at Chapel Hill, Chapel Hill, NC 27599, USA}
\affiliation{Department of Applied Physical Sciences, University of North Carolina at Chapel Hill, Chapel Hill, NC, 27599, USA}

\author{Rene Lopez}
\affiliation{Department of Physics and Astronomy, University of North Carolina at Chapel Hill, Chapel Hill, NC 27599, USA}
\affiliation{Department of Applied Physical Sciences, University of North Carolina at Chapel Hill, Chapel Hill, NC, 27599, USA}
\author{Wencan Jin}
\affiliation{Department of Physics, Auburn University, Auburn, AL 36849, USA}
\author{Yi Li}
\affiliation{Materials Science Division, Argonne National Laboratory, Argonne, IL 60439, USA}
\author{Binbin Yang}
\email{byang1@ncat.edu}
\affiliation{Department of Electrical and Computer Engineering, North Carolina A$\&$T State University, Greensboro, NC 27411, USA}
\author{Wei Zhang}
\email{zhwei@unc.edu}
\affiliation{Department of Physics and Astronomy, University of North Carolina at Chapel Hill, Chapel Hill, NC 27599, USA}

\date{\today}

\begin{abstract}
The spatial structure of a microwave photon mode provides a degree of freedom for controlling its interaction with a localized magnetic excitation. We demonstrate site-selective magnon--photon coupling in modular arrays of discrete dielectric resonators mounted on a common microstrip line. The resonators can be added, removed, or rearranged on the same completed PCB, enabling post-fabrication geometrical reconfiguration of the collective photon modes and their nodal structure. A YIG sphere is positioned successively above individual resonators, while an in-plane magnetic field tunes its Kittel mode through the collective photon resonances. In a three-resonator array, modes with finite local microwave magnetic fields hybridize with the magnon at every measured site, whereas a mode containing a central field node exhibits no resolvable splitting when the sphere is positioned at that node. Full-wave simulations and a coupled-mode model show that the interaction is governed by the local amplitude of the collective photon mode rather than by spectral resonance alone. Extending this local-node selection rule to a five-resonator array successfully predicts the measured position-dependent coupling patterns. These results establish post-fabrication control of spatial magnon--photon hybridization and provide a route toward reconfigurable magnonic interfaces in future hybrid quantum microwave architectures.
\end{abstract}

\maketitle

\section{Introduction}

Hybrid systems combining magnetic excitations with microwave photons provide a platform for studying coherent light--matter interactions and controlling information exchange between spin and electromagnetic degrees of freedom~\cite{Rameshti2022CavityMagnonics,Harder2021CoherentDissipative,awschalom2021quantum,zhang2026perspective,flebus20242024}. Yttrium iron garnet (YIG) is particularly suitable for cavity-magnonic experiments because of its high spin density, rich nonlinearity, and low magnetic damping~\cite{serga2010yig,Goryachev2014HighCooperativity,qu2025pump,wu2025coupling}. When the field-tunable Kittel mode~\cite{Kittel1948FerromagneticResonance} of a YIG sample is brought into resonance with a microwave photon mode, coherent coupling produces hybridized magnon--photon modes that appear as an avoided crossing in the transmission spectrum~\cite{Tabuchi2014Hybridizing,Zhang2014Strongly,Huebl2013HighCooperativity}. Most measurements identify and control this interaction through the spectral detuning between the magnon and photon resonances.

Spectral resonance, however, is only one requirement for magnon-photon hybridization. The effective coupling also depends on the spatial overlap and polarization matching between the magnon mode and the local microwave magnetic field~\cite{Bourhill2016Ultrahigh,xiong2024combinatorial,xiong2024hybrid,xiong2026photon,bourcin2024level}. Consequently, a magnetic sample placed at a field node can exhibit coupling suppression when the magnon and photon frequencies are degenerate~\cite{Harder2021CoherentDissipative,Gardin2024Engineering}. This spatial dependence becomes especially important in coupled-resonator arrays, where hybridization of the individual resonator modes produces collective photon modes with site-dependent amplitudes and phases~\cite{Pan2024Magnon}. The spatial eigenvectors of these modes can therefore provide a mechanism for selecting the resonator sites at which a localized magnetic excitation couples to a given collective photon mode~\cite{Wu2026Magnon,Hong2026Strong}.

A central feature of the present platform is that its collective-mode structure can be reconfigured without fabricating a new microwave circuit. In a conventional PCB resonator circuit, the resonant structures are defined by the patterned conductor geometry and remain fixed once the board has been fabricated. In contrast, the PCB in our platform provides the microstrip feedline, while the discrete dielectric resonators (DRs) can be added, removed, or rearranged on the same board. This post-fabrication geometrical reconfiguration changes the number of collective photon modes as well as their frequencies, spatial eigenvectors, and nodal sites. It therefore provides a direct means of changing where a localized magnetic excitation can couple to, or become decoupled from, a selected collective photon mode.

Here, we investigate site-selective magnon--photon coupling in modular arrays of dielectric resonators coupled through their near fields and excited by a common microstrip line. A millimeter-scale YIG sphere is moved among individual resonators, while an in-plane magnetic field tunes its Kittel mode through the collective photon resonances, enabling the interaction to be probed locally across each array. We find that different collective photon modes produce distinct spatial coupling patterns within the same resonator arrangement. Magnon--photon hybridization remains observable at sites with finite local microwave magnetic fields but is strongly suppressed when the YIG sphere is positioned at a collective-mode node. Full-wave simulations and a coupled-mode model consistently associate this site dependence with the spatial eigenvectors of the collective photon modes.

We first establish the connection between collective-mode nodes and site-selective magnon--photon coupling in a three-resonator array and then test its scalability in a five-resonator array with a richer nodal structure. Across both arrays, the measured position-dependent hybridization follows the node patterns of the collective photon modes, identifying their spatial structure as a control parameter complementary to spectral detuning. Because these mode profiles can be modified by adding, removing, or rearranging discrete resonators after the microwave board has been fabricated, the platform enables post-fabrication geometrical control over both the collective-mode spectrum and the spatial distribution of magnon--photon coupling. In low-loss implementations operated in the quantum regime, such spatial programmability could provide a building block for reconfigurable magnonic interfaces and selective interaction pathways in hybrid quantum microwave networks~\cite{Yuan2022QuantumMagnonics,Tabuchi2014Hybridizing,Morris2017StrongCoupling}.

\section{Experimental Setup and collective-mode formation}

\subsection{Experimental Configuration}

\begin{figure}[htbp]
    \centering
    \includegraphics[width=\columnwidth]{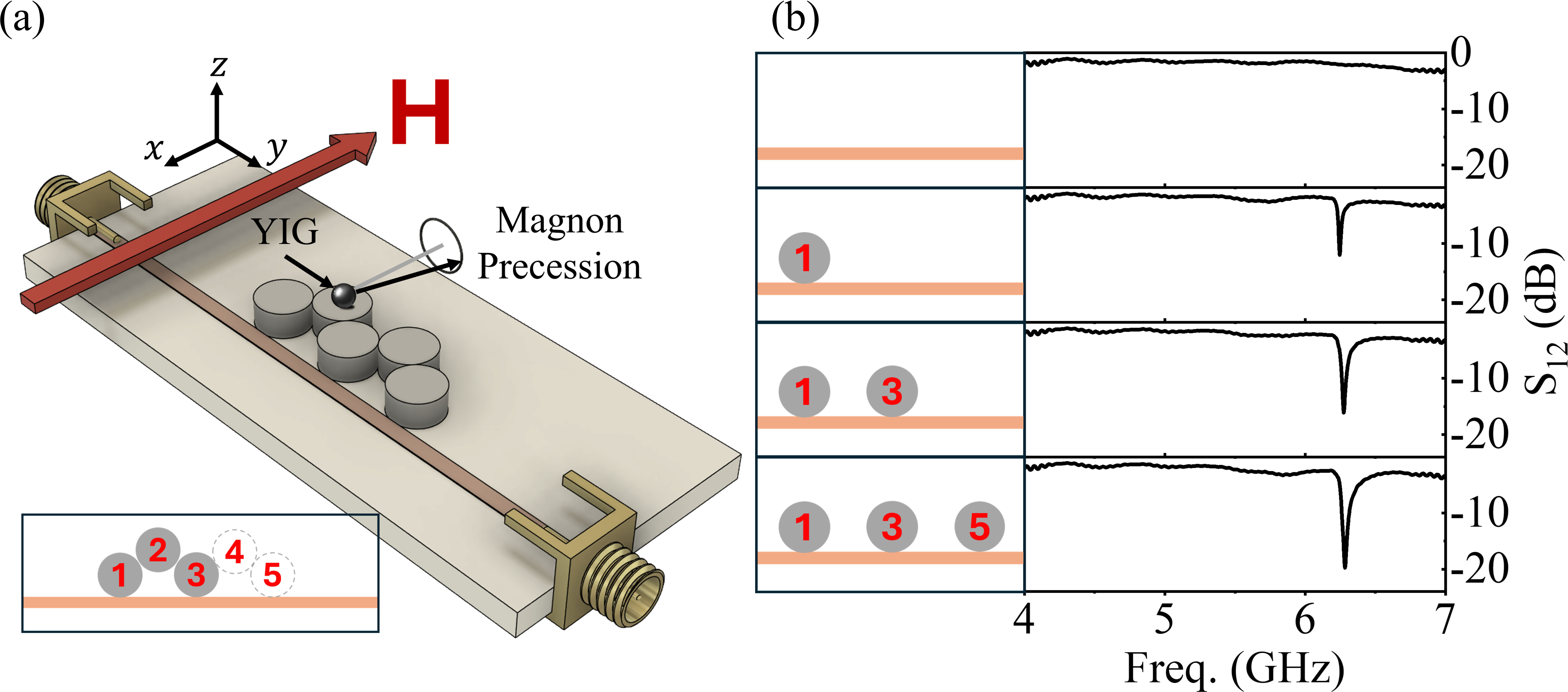}
    \caption{(a) Schematic of the zigzag DR array mounted along a copper microstrip line. A YIG sphere is placed directly above the selected DR. The DC bias field, $\mathbf{H}_\mathrm{ext}$,  is in the plane of the board. The inset defines the resonator indices $D_1$ to $D_5$: the filled resonators form the 3-DR array, whereas the dashed positions indicate the additional resonators reserved for constructing the 4-DR and 5-DR arrays. (b) $S_{12}$ spectra of multiple DRs separated by 6 mm. The absence of a resolvable normal-mode splitting indicates that any coupling-induced frequency separation at this distance remains smaller than the experimental linewidth.}
    \label{fig:1}
\end{figure}

In the experiment, the microstrip line is fabricated on a \(1.27~\mathrm{mm}\)-thick TMM10i PCB board from ROGERS Corporation. The substrate is cut to dimensions of $25 \times 100 \ \mathrm{mm^2}$, and a transmission line of 3.5 mm width is patterned by milling. Two ends of the transmission line are soldered to Sub-Miniature version A (SMA) connectors. The DRs are commercial DRT-6000 dielectric resonators with a relative permittivity of \(\epsilon_r=34\). They are mounted on the microstrip and labelled $D_1$ to $D_5$ according to their positions in the array, as shown in Fig.~\ref{fig:1}(a). The microstrip-excited DR geometry is similar to the zigzag-chain platform reported by Pan \textit{et al.}~\cite{Pan2024Magnon}, while the present work focuses on position-resolved magnon--photon coupling in reconfigurable finite DR arrays. Specifically, the filled circles in the inset define the 3-DR configuration, while the additional dashed circles are occupied when constructing the 4-DR and 5-DR arrays. A 1-mm-diameter YIG sphere is placed directly above the center of a selected DR. The PCB board with the YIG sphere is positioned horizontally inside the electromagnet so that the external magnetic field $\mathbf{H}_\mathrm{ext}$ is in the plane of the board and perpendicular to the transmission direction of the microwave. 

Scanning \(\mathbf{H}_{\mathrm{ext}}\) tunes the magnon resonance of the YIG sphere through the microwave photon-mode frequencies~\cite{Kittel1948FerromagneticResonance}. Repositioning the YIG sphere above different DRs enables the local magnon--photon interaction to be measured at different sites within the same resonator array. Both transmission coefficients, \(S_{12}\) and \(S_{21}\), are recorded using a Keysight E5071C vector network analyzer (VNA) connected to the two SMA ports of the PCB. Because the two transmission directions exhibit the same qualitative resonance and hybridization patterns, \(S_{12}\) is used consistently throughout this work. During measurement, \(\mathbf{H}_{\mathrm{ext}}\) is swept from \(-3000\) to \(+3000~\mathrm{Oe}\), while the transmission spectrum is acquired from \(100~\mathrm{kHz}\) to \(8.5~\mathrm{GHz}\). To isolate the magnetic-resonance response of the YIG sphere, we define the differential transmission magnitude in decibels as:
\begin{equation*}
\begin{aligned}
\Delta S_{12} &= S_{12}^\mathrm{raw}(H,f) - S_{12}^\mathrm{bg}(0,f), \\
S_{12} &= 20\log_{10}\left|S_{12}^{\mathrm{complex}}\right|.
\end{aligned}
\end{equation*}
where $S_{12}^\mathrm{raw}(H,f)$ is measured with the YIG sphere placed above the center of a selected DR under $\mathbf{H}_{ext}$, and $S_{12}^\mathrm{bg}(0,f)$ is the background spectrum acquired without the YIG sphere and under a zero bias field. 

We first construct a weak-coupling reference by placing DRs at locations 1, 3, and 5, corresponding to a nearest-neighbor separation of \(d=6~\mathrm{mm}\), as shown in Fig.~\ref{fig:1}(b). For configurations containing one, two, and three DRs, each spectrum retains a single resonance without a resolvable mode splitting, although the resonance depth and linewidth both increase as additional DRs are introduced. This observation indicates that any coupling-induced frequency separation at \(d=6~\mathrm{mm}\) remains smaller than the measured linewidth. We therefore treat this configuration as a weak-coupling reference rather than a strictly uncoupled limit. Distance-dependent measurements of a two-DR system that support the choice of \(d=6~\mathrm{mm}\) as the weak-coupling reference are presented in Appendix~\ref{app:distance_dependence}. This reference distinguishes the response of a single DR from the collective modes that emerge when the resonators are brought close enough.


\subsection{Formation of collective photon modes in the 3-DR array}

Dielectric microwave resonators support spatially confined electromagnetic modes, while coupling between resonant structures redistributes both their eigenfrequencies and field profiles~\cite{Mett2008DielectricMicrowave,Elnaggar2014CoupledModes}. In our system, we realize this coupled-mode regime by placing three DRs at locations 1, 2, and 3, where adjacent resonators interact predominantly through electric-dipolar near-field coupling. This interaction lifts the degeneracy of the isolated-DR resonance and produces collective photon modes whose electromagnetic fields are distributed over the three resonators~\cite{Pan2024Magnon}.

\begin{figure}[htbp]
    \centering
    \includegraphics[width=\columnwidth]{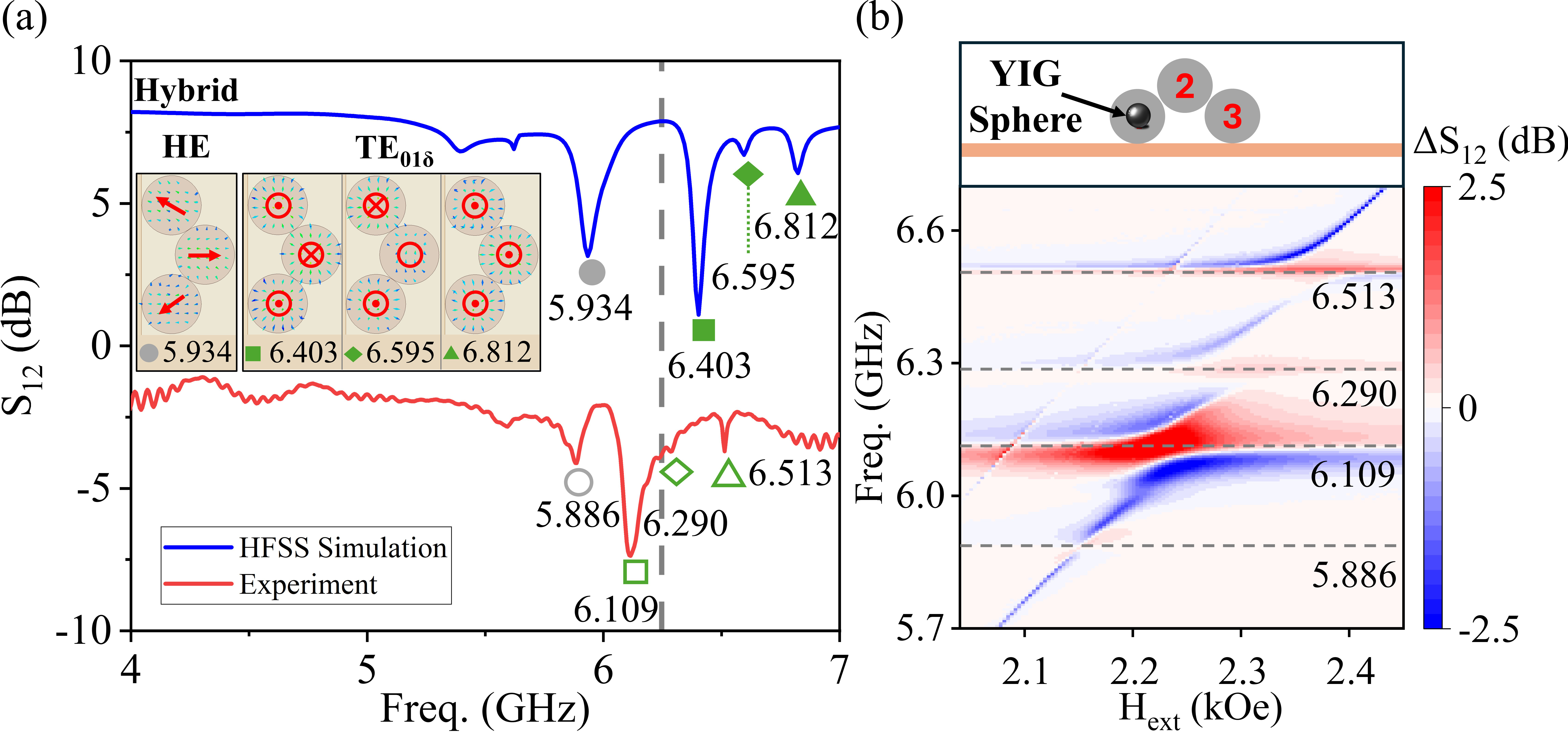}
    \caption{
    (a) Measured zero-field transmission \(S_{12}\) of the 3-DR array (red) and the corresponding HFSS spectrum (blue, shifted vertically by \(+8.5~\mathrm{dB}\) for clarity). The vertical dashed line marks the single-DR resonance, while solid and hollow symbols identify the simulated and experimental resonances, respectively, with matching shapes indicating the proposed mode correspondence. The insets show the HFSS microwave magnetic-field profiles at \(5.934\), \(6.403\), \(6.595\), and \(6.812~\mathrm{GHz}\), evaluated at a common RF phase. The laboratory coordinate system is defined in Fig.~\ref{fig:1}(a).
    (b) Differential transmission magnitude \(\Delta S_{12}\) as a function of external magnetic field and frequency, measured with the YIG sphere positioned above \(D_1\), as indicated by the black sphere in the upper schematic. The labeled horizontal resonances identify the experimentally observed photon-mode frequencies. 
    }
    \label{fig:2}
\end{figure}

In Fig.~\ref{fig:2}(a), we compare the measured zero-field transmission magnitude \(S_{12}\) with the corresponding full-wave simulation performed using Ansys HFSS. The vertical dashed line marks the resonance frequency of an isolated DR and provides a reference for the coupling-induced redistribution of the resonances. Solid and hollow symbols denote the simulated and experimentally identified resonances, respectively, with matching symbol shapes indicating the proposed correspondence between them. 

The insets in Fig.~\ref{fig:2}(a) show the HFSS-calculated microwave magnetic-field profiles of the four labeled resonances, evaluated at a common RF phase. The lower-frequency resonance at \(5.934~\mathrm{GHz}\) is classified as a hybrid electromagnetic (HEM) mode~\cite{luk2003dielectric}, for which both longitudinal electric- and magnetic-field components are generally nonzero. By contrast, the three higher-frequency resonances at \(6.403\), \(6.595\), and \(6.812~\mathrm{GHz}\) originate from the \(\mathrm{TE}_{01\delta}\) mode of an individual DR. This mode is characterized by a predominantly axial microwave magnetic field and an azimuthal electric field. The \(\mathrm{TE}_{01\delta}\) notation denotes an azimuthally uniform, first-radial-order mode with finite axial confinement. The arrows indicate the dominant in-plane microwave magnetic-field directions. The \(\odot\) and \(\otimes\) symbols denote local fields directed along \(+z\) and \(-z\), respectively, while \(\scalebox{0.8}{$\bigcirc$}\) marks a node of the \(z\)-directed component. The three \(\mathrm{TE}_{01\delta}\)-derived modes constitute the collective-mode manifold considered in this work and are associated with the experimentally observed resonances at \(6.109\), \(6.290\), and \(6.513~\mathrm{GHz}\), respectively. This mode correspondence is examined more directly in Fig.~\ref{fig:3} by comparing the simulated spatial field profiles with the position-dependent magnon--photon hybridization.

To probe the magnon-photon interaction, we place the YIG sphere at the top of the resonator $D_1$ and scan the external field, as illustrated in Fig.~\ref{fig:2}(b). As \(\mathbf{H}_{\mathrm{ext}}\) is swept, the field-dispersive Kittel-mode branch crosses several nearly field-independent photon resonances, producing characteristic magnon--photon hybridization features~\cite{Harder2021CoherentDissipative}. All three \(\mathrm{TE}_{01\delta}\)-derived modes identified in Fig.~\ref{fig:2}(a) can be tracked in the field-dependent map. In particular, the mode near \(6.290~\mathrm{GHz}\), although only weakly visible in the zero-field transmission spectrum, exhibits a resolvable avoided crossing when the YIG sphere is positioned above \(D_1\). This demonstrates that the small direct-transmission response of this mode does not imply a vanishing internal microwave field or a vanishing local magnon--photon coupling.

An additional experimental resonance appears near \(5.886~\mathrm{GHz}\), close to the simulated hybrid HEM mode at \(5.934~\mathrm{GHz}\). Unlike the three higher-frequency modes, this lower-frequency feature does not originate from the \(\mathrm{TE}_{01\delta}\)-derived collective-mode manifold and does not exhibit a resolvable avoided crossing for the YIG sphere above \(D_1\). This weak coupling also originates from the fact that the HEM mode displays a horizontal RF magnetic field, which couples weakly to an in-plane biased magnon mode. Because the following coupled-mode model uses one local \(\mathrm{TE}_{01\delta}\)-like photon mode per DR, the lower-frequency hybrid mode is not included in the three-mode analysis. We therefore focus on the collective photon modes near \(6.109\), \(6.290\), and \(6.513~\mathrm{GHz}\) in this work.

\section{Magnon--photon coupling in zigzag resonator arrays}

\subsection{Position-dependent coupling in the three-resonator array}

\begin{figure*}[htb]
    \centering
    \includegraphics[width= 6 in]{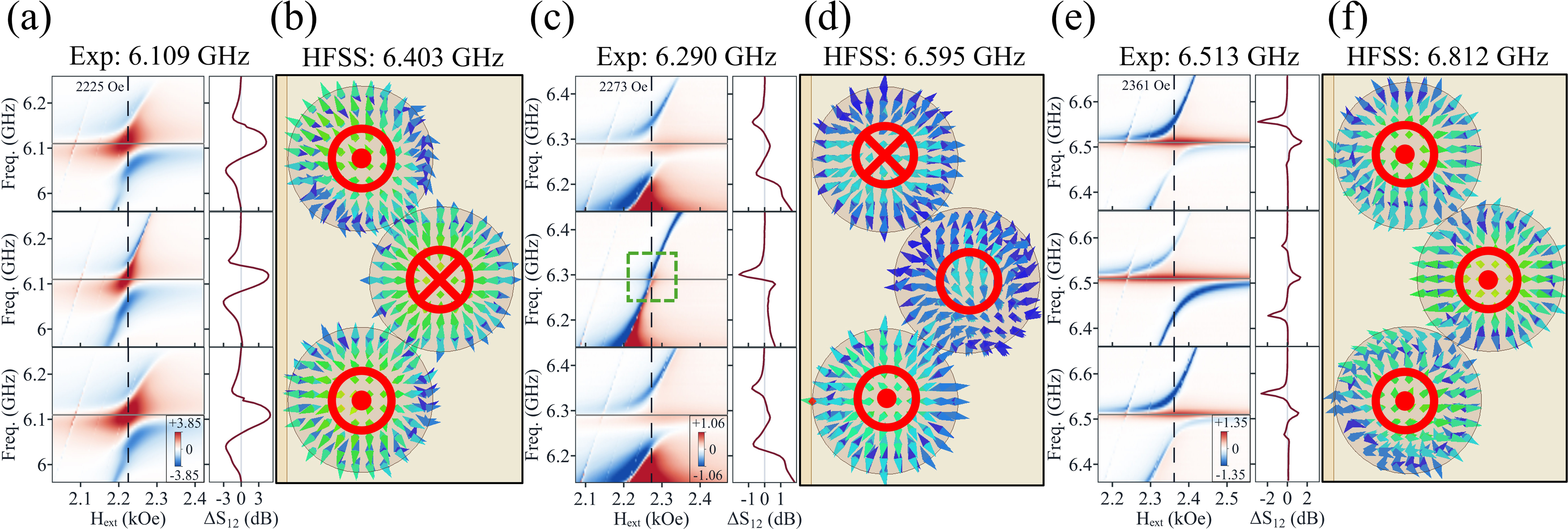}
    \caption{Local-node selection rule in the 3-DR array. The experimental differential transmission-magnitude maps \(\Delta S_{12}\) at (a) \(6.109\), (c) \(6.290\), and (e) \(6.513~\mathrm{GHz}\) are compared with the corresponding HFSS microwave magnetic-field profiles at (b) \(6.403\), (d) \(6.595\), and (f) \(6.812~\mathrm{GHz}\), respectively. Within each experimental panel, the top, middle, and bottom rows correspond to the YIG sphere positioned directly above \(D_1\), \(D_2\), and \(D_3\), as indicated by the accompanying schematics. The spectra extracted at the magnetic fields marked by the vertical dashed lines are displayed beside the corresponding maps. The green dashed box in (c) marks the site--mode combination for which no resolvable avoided crossing is observed. The \(\odot\), \(\otimes\), and \scalebox{0.8}{$\bigcirc$} symbols denote local microwave magnetic fields directed along \(+z\), along \(-z\), and a magnetic-field node, respectively. The laboratory coordinate system is defined in Fig.~\ref{fig:1}(a).}
    \label{fig:3}
\end{figure*}

Having identified the collective photon modes of the 3-DR array, we next investigate how their interaction with the YIG sphere depends on the position of the sphere. The YIG sphere is sequentially positioned directly above \(D_1\), \(D_2\), and \(D_3\), and the external magnetic field is swept through the corresponding magnon--photon resonance conditions. Figures~\ref{fig:3}(a), (c), and (e) show the position-resolved measurements, while Figs.~\ref{fig:3}(b), (d), and (f) show the corresponding HFSS-calculated microwave magnetic-field profiles. The experiment--simulation mode assignments used in the following analysis are
\[
\begin{aligned}
6.109~\mathrm{GHz}\;(\mathrm{experiment})
&\leftrightarrow
6.403~\mathrm{GHz}\;(\mathrm{HFSS}),\\
6.290~\mathrm{GHz}\;(\mathrm{experiment})
&\leftrightarrow
6.595~\mathrm{GHz}\;(\mathrm{HFSS}),\\
6.513~\mathrm{GHz}\;(\mathrm{experiment})
&\leftrightarrow
6.812~\mathrm{GHz}\;(\mathrm{HFSS}).
\end{aligned}
\]

For the collective mode near 6.109 GHz [Fig.~\ref{fig:3}(a)], a resolvable avoided crossing is observed at all three YIG positions. The magnon Kittel mode branch hybridizes with the nearly field-independent photon resonance, producing two separated hybridized branches near resonance. Similar avoided crossings are observed for the 6.513 GHz mode at all three positions [Fig.~\ref{fig:3}(e)]. These measurements show that the YIG sphere couples to both the lowest- and highest-frequency modes of the selected \(\mathrm{TE}_{01\delta}\)-derived collective-mode manifold at each of the three resonator sites. A qualitatively different position dependence is observed for the mode near 6.290 GHz in Fig.~\ref{fig:3}(c). When the YIG sphere is positioned above $D_1$ or $D_3$, the transmission maps show an avoided crossing between the magnon and photon branches. In contrast, when the sphere is positioned above $D_2$, the two branches cross without a resolvable splitting. This suppressed-coupling case is highlighted by the green dashed box. These measurements therefore identify the \(6.290\)-GHz resonance as the middle-frequency mode of the selected \(\mathrm{TE}_{01\delta}\)-derived collective-mode manifold, characterized by finite magnon--photon coupling at the two outer resonators and strongly suppressed coupling at the central resonator.

The spectra extracted at the magnetic fields indicated by the vertical dashed lines are shown to the right of the corresponding maps and provide a direct comparison of the hybridization at the three YIG positions. For the 6.109 and 6.513 GHz modes, the vertical cuts exhibit a resolved
two-resonance response at all three sites. For the 6.290 GHz mode, a split response is observed with the YIG sphere above $D_1$ and $D_3$, whereas a single unsplit resonance is detected above $D_2$. The cut spectra therefore corroborate the position dependence observed in the two-dimensional transmission maps and identify $D_2$ as a suppressed-coupling site for the 6.290 GHz collective mode.

The simulated microwave magnetic-field profiles provide the spatial information needed to interpret these measurements. The \(6.595\)-GHz HFSS mode assigned to the experimental \(6.290\)-GHz resonance exhibits a magnetic-field node at the middle resonator \(D_2\), while retaining finite local field amplitudes at \(D_1\) and \(D_3\) [Fig.~\ref{fig:3}(d)]. This profile is consistent with the experimentally observed avoided crossings above \(D_1\) and \(D_3\) and the absence of a resolvable splitting above \(D_2\). By contrast, the simulated modes at \(6.403\) and \(6.812~\mathrm{GHz}\) have finite local microwave magnetic-field amplitudes at all three resonators [Figs.~\ref{fig:3}(b) and (f)], consistent with the avoided crossings observed at every measured YIG position for the corresponding experimental modes at \(6.109\) and \(6.513~\mathrm{GHz}\).

Taken together, the mode ordering and the agreement between the simulated spatial profiles and the measured site-dependent hybridization support the three proposed experiment--simulation mode assignments. More generally, these results show that spectral resonance between the magnon and photon modes is not sufficient by itself to determine the interaction strength. The effective coupling also depends on the local amplitude of the selected collective photon mode at the YIG position. In particular, a collective-mode node at the occupied resonator suppresses the interaction even when the magnon and photon frequencies are tuned into resonance. This observation motivates a local coupling selection rule, which is derived for the three-resonator system in the following subsection.

\subsection{Three-resonator model and local coupling selection rule}

To account for the observed site-dependent coupling, we model the 3-DR array as a chain of nominally identical local \(\mathrm{TE}_{01\delta}\)-like photon modes with dominant nearest-neighbor coupling. Let \(\hat c_i\) annihilate a photon in the local mode of resonator \(D_i\), where \(i=1,2,3\). Neglecting the direct coupling between \(D_1\) and \(D_3\), the photon Hamiltonian is: 
\begin{equation}
    \frac{\hat H_{\mathrm{ph}}^{(3)}}{\hbar}
    =\omega_0\sum_{i=1}^{3}\hat c_i^\dagger\hat c_i
    +J\sum_{i=1}^{2}
    \left(\hat c_i^\dagger\hat c_{i+1}
    +\hat c_{i+1}^\dagger\hat c_i\right).
\label{eq:three_dr_photon_hamiltonian}
\end{equation}
Diagonalizing Eq.~\ref{eq:three_dr_photon_hamiltonian} gives three collective-mode operators \(\hat a_n=\sum_i v_{n,i}\hat c_i\), where \(v_{n,i}\) is the amplitude of the local photon mode, and the corresponding eigenfrequencies \(\Omega_n\): 
\begin{equation}
\begin{aligned}
    \hat{a}_1 &= \frac{\hat{c}_1-\sqrt{2}\hat{c}_2+\hat{c}_3}{2},
    &\qquad
    \Omega_1 &= \omega_0-\sqrt{2}J, \\
    \hat{a}_2 &= \frac{\hat{c}_1-\hat{c}_3}{\sqrt{2}},
    &\qquad
    \Omega_2 &= \omega_0, \\
    \hat{a}_3 &= \frac{\hat{c}_1+\sqrt{2}\hat{c}_2+\hat{c}_3}{2},
    &\qquad
    \Omega_3 &= \omega_0+\sqrt{2}J.
\end{aligned}
\label{eq:collective_mode_operators}
\end{equation}
For a measurement with the YIG sphere positioned above \(D_j\), we denote its Kittel-mode annihilation operator by \(\hat m_j\), with the field-tunable frequency \(\omega_m(H_{\mathrm{ext}})\). The YIG sphere couples predominantly to the local microwave field of the resonator directly beneath it. Within this local approximation and the rotating-wave approximation, the interaction Hamiltonian is~\cite{Rameshti2022CavityMagnonics}:
\begin{equation}
\begin{aligned}
    \frac{\hat H_{\mathrm{int}}^{(D_j)}}{\hbar}
    &=g_j\hat c_j^\dagger\hat m_j
    +g_j^{*}\hat c_j\hat m_j^\dagger\\
    &=\sum_{n=1}^{3}
    \left(G_{n,D_j}\hat a_n^\dagger\hat m_j
    +G_{n,D_j}^{*}\hat a_n\hat m_j^\dagger\right)  
\end{aligned}
\label{eq:magnon_hamiltonian}
\end{equation}
with
\begin{equation}
    G_{n,D_j}=g_j v_{n,j},
\label{eq:G_n_definition}
\end{equation}
where \(g_j\) is the local-basis magnon--photon coupling rate between the Kittel mode and the photon mode \(\hat c_j\) associated with \(D_j\)~\cite{Hong2026Strong}. Including the Kittel mode, the Hamiltonian of the three collective photon modes and the YIG sphere positioned above \(D_j\) can be written as
\begin{equation}
\begin{aligned}
    \frac{\hat H^{(D_j)}}{\hbar}
    =&
    \sum_{n=1}^{3}\Omega_n\hat a_n^\dagger\hat a_n
    +\omega_m(H_{\mathrm{ext}})\hat m_j^\dagger\hat m_j\\
    &+\sum_{n=1}^{3}
    \left(
    G_{n,D_j}\hat a_n^\dagger\hat m_j
    +G_{n,D_j}^{*}\hat a_n\hat m_j^\dagger
    \right),
\end{aligned}
\label{eq:full_collective_hamiltonian}
\end{equation}
When the Kittel mode is tuned near a selected collective photon mode \(n\) and the other two photon modes remain sufficiently far detuned, the full Hamiltonian can be reduced to the corresponding two-mode subspace. Within the ideal lossless approximation, the hybridized eigenfrequencies are
\begin{equation}
    \omega_{\pm}
    =
    \frac{\Omega_n+\omega_m(H_{\mathrm{ext}})}{2}
    \pm
    \sqrt{
    \left[\frac{\Omega_n-\omega_m(H_{\mathrm{ext}})}{2}\right]^2
    +\left|G_{n,D_j}\right|^2
    }.
\label{eq:hybridized_eigenfrequencies}
\end{equation}
At resonance, \(\omega_m(H_{\mathrm{ext}})=\Omega_n\), the ideal lossless model gives a mode separation of \(2|G_{n,D_j}|\). Thus, \(|G_{n,D_j}|\) determines the avoided-crossing gap measured with the YIG sphere above \(D_j\). If \(v_{n,j}=0\), then \(G_{n,D_j}=0\), and the Hamiltonian predicts a crossing rather than an avoided crossing.

Using \(G_{n,D_j}=g_jv_{n,j}\), the effective coupling rates of all three collective photon modes at the three resonator sites \(D_j\) can be calculated from the collective photon mode amplitudes \(v_{n,j}\):
\begin{equation}
\begin{pmatrix}
G_{1,D_1} & G_{1,D_2} & G_{1,D_3}\\[6pt]
G_{2,D_1} & G_{2,D_2} & G_{2,D_3}\\[6pt]
G_{3,D_1} & G_{3,D_2} & G_{3,D_3}
\end{pmatrix}
=
\begin{pmatrix}
\dfrac{g_1}{2} & -\dfrac{g_2}{\sqrt{2}} & \dfrac{g_3}{2}\\[10pt]
\dfrac{g_1}{\sqrt{2}} & 0 & -\dfrac{g_3}{\sqrt{2}}\\[10pt]
\dfrac{g_1}{2} & \dfrac{g_2}{\sqrt{2}} & \dfrac{g_3}{2}
\end{pmatrix},
\label{eq:all_G_3DR}
\end{equation}
where the rows correspond to the collective modes \(n=1,2,3\), and the columns correspond to the YIG positions \(D_1,D_2,D_3\). For compactness, we define the effective coupling vector of collective mode \(n\) as \(\mathbf G_n^{(3)}\equiv(G_{n,D_1},G_{n,D_2},G_{n,D_3})\). With the phases of the local photon modes chosen such that \(g_j\) is real and positive, the relative coupling-phase and node patterns are therefore
\begin{equation}
\begin{aligned}
\mathbf G_1^{(3)} &\sim (+,-,+),\\
\mathbf G_2^{(3)} &\sim (+,0,-),\\
\mathbf G_3^{(3)} &\sim (+,+,+).
\end{aligned}
\label{eq:G_sign_patterns}
\end{equation}

These analytical patterns are in qualitative agreement with the HFSS-calculated microwave magnetic-field profiles. At a common RF phase, the simulated modes at \(6.403\), \(6.595\), and \(6.812~\mathrm{GHz}\) exhibit the field-direction patterns \(\bigl(\odot,\otimes,\odot\bigr)\), \(\bigl(\otimes,\scalebox{0.8}{$\bigcirc$},\odot\bigr)\), and \(\bigl(\odot,\odot,\odot\bigr)\), respectively [Figs.~\ref{fig:3}(b), (d), and (f)]. The second HFSS pattern differs from \(\mathbf G_2^{(3)}\sim(+,0,-)\) only by an overall sign, which reflects the arbitrary global phase of the eigenmode and has no physical consequence. Moreover, the relative signs of the nonzero coupling rates determine the local coupling phases but do not change the avoided-crossing gap \(2|G_{n,D_j}|\).

Equation~\ref{eq:all_G_3DR} reproduces the observed site-dependent selection rule: the ideal-model result \(G_{2,D_2}=0\) predicts a crossing for the second collective mode at \(D_2\), consistent with the absence of a resolvable avoided crossing for the experimental \(6.290\)-GHz mode in Fig.~\ref{fig:3}(c), whereas the remaining nonzero matrix elements are consistent with the avoided crossings observed at the other mode--position combinations. The agreement among the analytical coupling patterns, the HFSS field-direction profiles, and the position-resolved measurements establishes the local collective-mode amplitude as the origin of the observed site-dependent magnon--photon interaction.

\subsection{Five-resonator validation of the local-node selection rule}

The three-resonator results establish that the interaction between the YIG sphere and a collective photon mode is suppressed when the sphere is positioned above a node of the local microwave magnetic field. To test whether this local-node selection rule remains valid in a larger and spatially more complex system, we next examine a five-resonator array.

\begin{figure}[htbp]
    \centering
    \includegraphics[width=2.7 in]{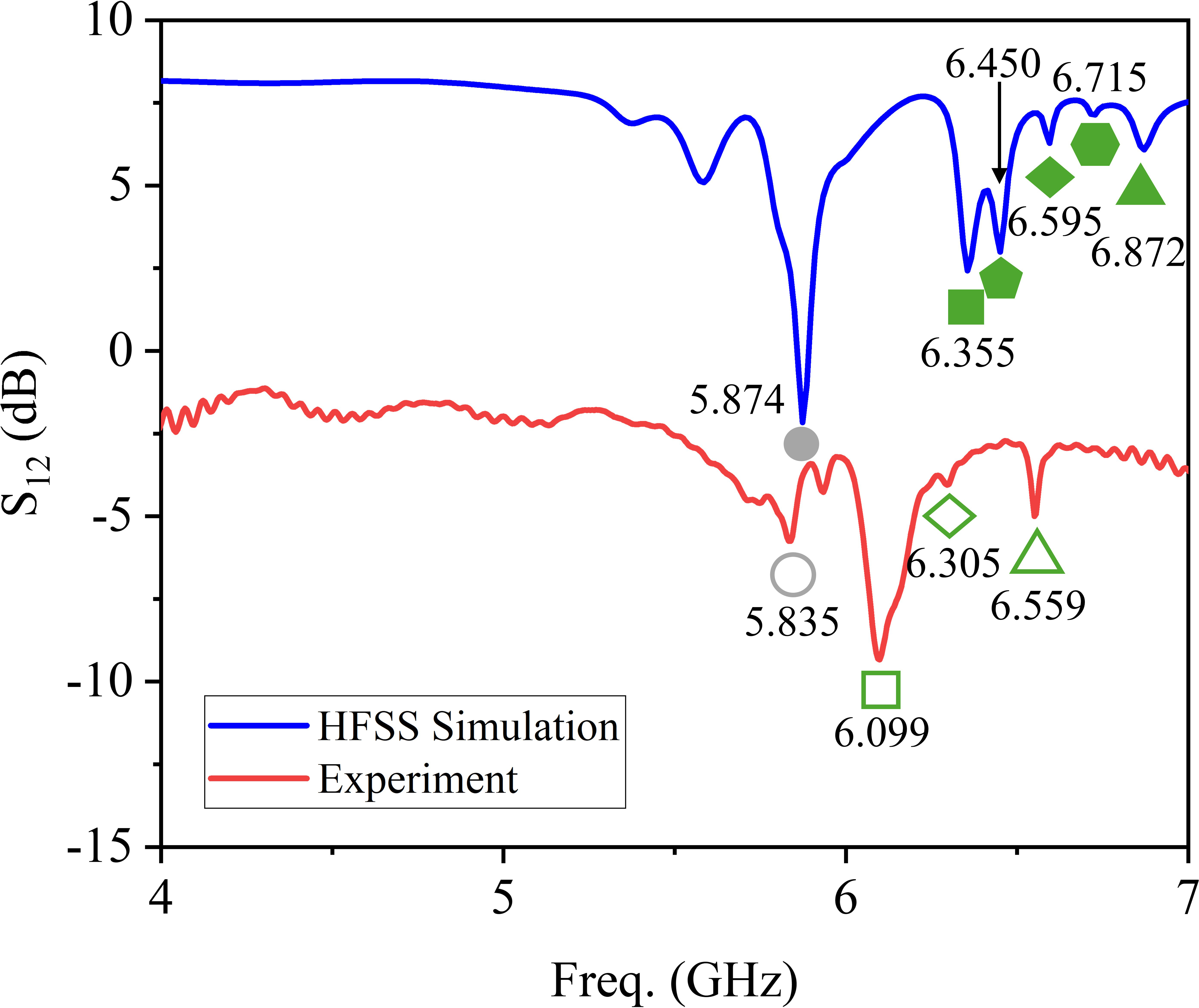}
    \caption{Measured zero-field transmission \(S_{12}\) of the compact 5-DR array (red), compared with the corresponding HFSS driven-response spectrum (blue). The measurement is performed without the YIG sphere and without an applied bias field, and the simulated spectrum is shifted vertically by \(+8.5~\mathrm{dB}\) for clarity. Solid and hollow symbols identify the simulated and experimental collective photon resonances, respectively, with matching shapes indicating the proposed mode correspondence.}
    \label{fig:4}
\end{figure}

We first identify the collective photon modes used in the position-resolved measurements. Fig.~\ref{fig:4} compares the measured zero-field transmission \(S_{12}\) of the compact 5-DR array with the corresponding HFSS driven-response spectrum. The five selected experimental resonances, ordered from low to high, occur at approximately \(6.095\), \(6.172\), \(6.305\), \(6.400\), and \(6.559~\mathrm{GHz}\). They are associated with the HFSS-calculated modes at \(6.355\), \(6.450\), \(6.595\), \(6.715\), and \(6.872~\mathrm{GHz}\), respectively, according to their frequency ordering. Matching symbols indicate the proposed experiment--simulation mode correspondence, which is further tested below through the simulated spatial field profiles and the measured position-dependent hybridization. Additional lower-frequency resonances lie outside the selected \(\mathrm{TE}_{01\delta}\)-derived collective-mode manifold and are not included in the following coupling analysis.

\begin{figure*}[htb]
    \centering
    \includegraphics[width=5.3 in]{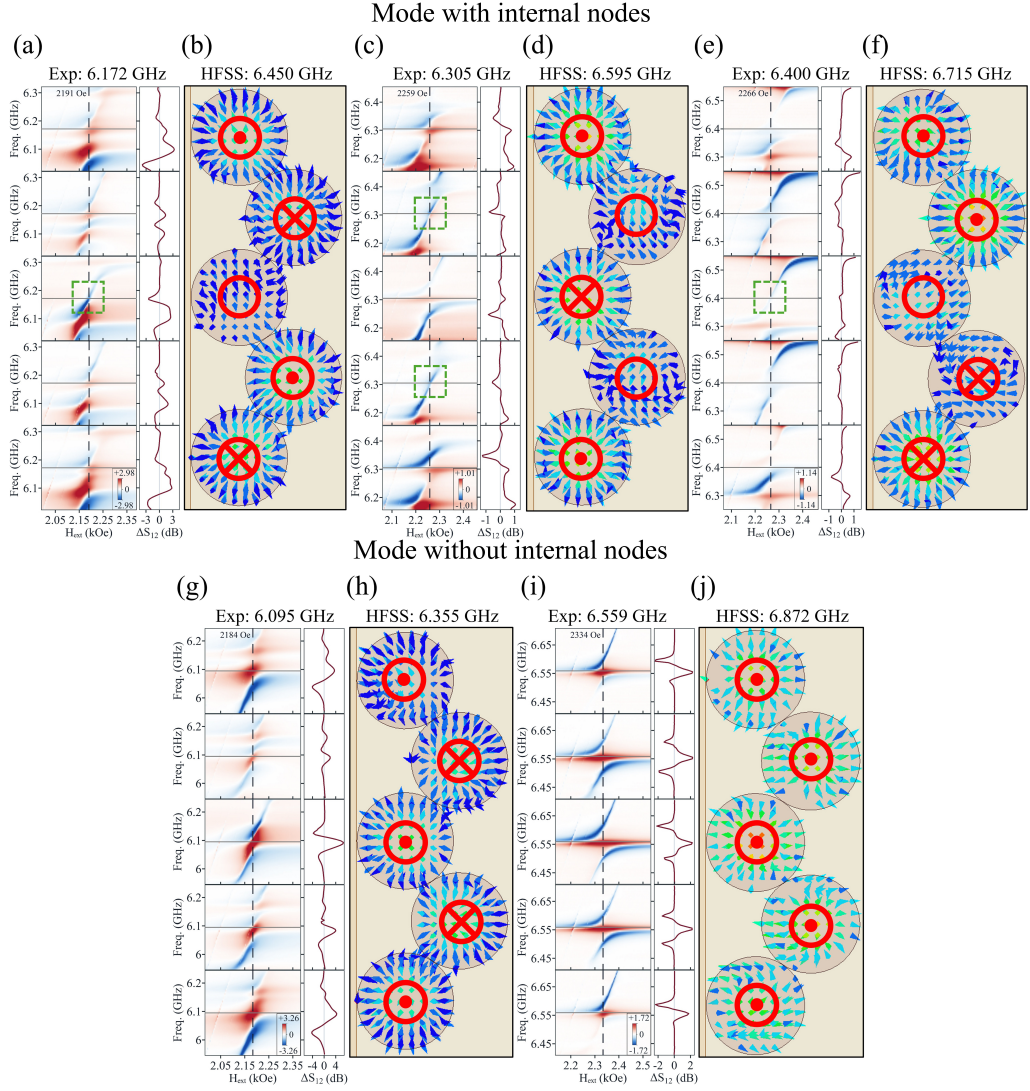}
    \caption{Local-node selection rule in the 5-DR array. The upper group shows modes with internal nodes, comparing the experimental \(\Delta S_{12}\) maps at (a) \(6.172\), (c) \(6.305\), and (e) \(6.400~\mathrm{GHz}\) with the corresponding HFSS microwave magnetic-field profiles at (b) \(6.450\), (d) \(6.595\), and (f) \(6.715~\mathrm{GHz}\), respectively. The lower group shows modes without internal nodes, comparing the experimental maps at (g) \(6.095\) and (i) \(6.559~\mathrm{GHz}\) with the corresponding HFSS profiles at (h) \(6.355\) and (j) \(6.872~\mathrm{GHz}\), respectively. Within each experimental panel, the five rows from top to bottom correspond to the YIG sphere positioned directly above \(D_1\) through \(D_5\). The solid horizontal lines mark the collective photon-mode frequencies, and the vertical dashed lines indicate the magnetic fields at which the spectra displayed beside the maps are extracted. Green dashed boxes mark site--mode combinations for which no resolvable avoided crossing is observed. The \(\odot\), \(\otimes\), and \(\scalebox{0.8}{$\bigcirc$}\) symbols in the HFSS panels denote local microwave magnetic fields directed along \(+z\), along \(-z\), and a magnetic-field node, respectively.}
\label{fig:5}
\end{figure*}

With the five selected modes indexed in ascending order of frequency, the uniform nearest-neighbor model gives the following relative phase and node patterns of the effective coupling vectors, using the same phase convention as in the three-resonator model:
\begin{equation}
\begin{aligned}
\mathbf G_1^{(5)}&\sim(+,-,+,-,+),\\
\mathbf G_2^{(5)}&\sim(+,-,0,+,-),\\
\mathbf G_3^{(5)}&\sim(+,0,-,0,+),\\
\mathbf G_4^{(5)}&\sim(+,+,0,-,-),\\
\mathbf G_5^{(5)}&\sim(+,+,+,+,+).
\end{aligned}
\label{eq:five_dr_sign_patterns}
\end{equation}
The signs specify the relative phases of the effective local coupling rates, while the zero components originate from collective-mode nodes and identify the sites at which the ideal model predicts suppressed coupling. Using the mode assignments in Fig.~\ref{fig:4}, the predicted nodes occur above \(D_3\) for the experimental \(6.172\)- and \(6.400\)-GHz modes and above \(D_2\) and \(D_4\) for the \(6.305\)-GHz mode, whereas no nodal suppression is predicted for the \(6.095\)- and \(6.559\)-GHz modes.

Fig.~\ref{fig:5} tests these predictions using position-resolved measurements of the 5-DR array, with the modes grouped according to their predicted node structures. Within each experimental panel, the five rows correspond from top to bottom to the YIG sphere positioned above \(D_1\) through \(D_5\). For the three modes with internal nodes, no resolvable avoided crossing is observed above \(D_3\) for the experimental modes at \(6.172\) and \(6.400~\mathrm{GHz}\) [Figs.~\ref{fig:5}(a) and (e)], or above \(D_2\) and \(D_4\) for the \(6.305\)-GHz mode [Fig.~\ref{fig:5}(c)]. These suppressed-coupling sites coincide with the nodes of the \(z\)-directed microwave magnetic-field component in the corresponding HFSS profiles at \(6.450\), \(6.595\), and \(6.715~\mathrm{GHz}\) [Figs.~\ref{fig:5}(b), (d), and (f)] and with the node positions predicted by Eq.~\ref{eq:five_dr_sign_patterns}. Resolvable avoided crossings are observed at the remaining resonator sites, consistent with their finite local microwave magnetic fields. The two modes without internal nodes provide a complementary control. For the experimental modes near \(6.095\) and \(6.559~\mathrm{GHz}\) [Figs.~\ref{fig:5}(g) and (i)], resolvable avoided crossings are observed at all five YIG positions. Their corresponding HFSS profiles at \(6.355\) and \(6.872~\mathrm{GHz}\) have finite local microwave magnetic fields at every resonator [Figs.~\ref{fig:5}(h) and (j)], consistent with the absence of internal nodes in the five-resonator model.

The agreement among the five-resonator model, the HFSS-calculated field profiles, and the measured position-dependent avoided crossings shows that the local-node selection rule established in the three-resonator array remains valid in the larger five-resonator system. The node structure identifies the site--mode combinations for which the interaction is suppressed, while the coupling strengths at the remaining sites also depend on the local magnon--photon overlap. These results demonstrate that spectral resonance alone is insufficient to determine the interaction; its site dependence is governed by the spatial structure of the collective photon mode. This node-based description is not restricted to the specific array sizes examined here, and its extension to an arbitrary number of resonators is derived in Appendix~\ref{app:N_DR_generalization}.

\section{Discussion and conclusion}

The position-resolved measurements demonstrate that spectral resonance alone does not determine the magnon--photon interaction in a coupled-resonator array. Across both the three- and five-resonator systems, hybridization is suppressed at the nodal sites predicted by the collective-mode eigenvectors, while resolvable avoided crossings remain at sites with finite local microwave magnetic fields. This agreement shows that the YIG sphere acts as a local probe of the collective photon-mode structure and that the local modal amplitude provides a spatial control parameter complementary to spectral detuning. This interpretation is consistent with recent photonic-lattice studies in which the coupling of a localized YIG sphere follows the local photonic eigenmode amplitude~\cite{Hong2026Strong,Wu2026Magnon}.

The use of discrete dielectric resonators provides a modular approach for controlling these spatial eigenvectors. Reconfigurable magnon--photon interactions can also be realized by controlling the magnetic subsystem. For example, rotating the bias field relative to a spatially nonuniform microwave field enables in-situ tuning of the coherent coupling, magnon damping, and nonreciprocity~\cite{Shuai2025Reconfigurable}. The present approach is complementary: it reconfigures the photonic subsystem by rearranging discrete dielectric resonators after the PCB has been fabricated. Whereas the resonant structures in a conventional PCB resonator circuit are fixed by the patterned conductor geometry, the PCB in our platform provides only the microstrip feedline, while the discrete DR elements can be added, removed, or rearranged without refabricating the board. Such post-fabrication geometrical reconfiguration modifies the collective-mode frequencies and nodal structure and thereby changes the resonator sites at which magnon--photon coupling is allowed or suppressed. More generally, microwave-resonator geometry provides an effective means of reshaping the photonic mode spectrum and enhancing magnon--photon coupling~\cite{xiong2024combinatorial}.

The uniform nearest-neighbor model captures the node patterns observed in the 3-DR and 5-DR arrays but remains an idealized description of the experimental system. The 4-DR results reveal a deviation from this model, while the agreement between the measurements and full-wave simulations is consistent with geometry-dependent microwave-field distributions not captured by the uniform-chain approximation. Thus, the collective-mode node structure provides the leading-order selection rule, whereas the specific resonator and feedline geometry can further modify the local magnon--photon interaction. A detailed comparison is presented in Appendix~\ref{app:4DR_map}.

In conclusion, modular DR arrays support collective photon modes with spatially structured eigenvectors and predictable nodal sites. Projecting the local YIG--photon interaction onto these collective eigenmodes produces a site-dependent coupling selection rule: a finite local mode amplitude permits magnon--photon hybridization, whereas a node suppresses the interaction within the experimental resolution. The selection rule established in the three-resonator system predicts the coupling pattern measured in the five-resonator array, demonstrating a scalable connection between collective-mode structure and local magnon--photon coupling. This connection provides a basis for extending the present approach to larger or deliberately nonuniform resonator networks, in which multiple collective photon modes could mediate spatially designed interactions among several magnetic elements. The feasibility of scaling cavity-magnonic systems beyond a single magnetic element has been demonstrated using multiple ferrimagnetic spheres coupled to a common microwave cavity~\cite{Crescini2021Coherent}. Incorporating additional degrees of freedom, such as mechanical modes coupled through magnetostriction~\cite{Zhang2016CavityMagnomechanics}, could further extend this platform toward multimode hybrid interfaces~\cite{xiong2024magnon,zhu2026strong,christy2026modulating,christy2025tuning,zhou2026adaptive}. With low-loss resonators and operation in the quantum regime, geometrically reconfigurable coupling architectures may ultimately provide programmable magnonic interfaces and selective interaction pathways for hybrid quantum microwave networks~\cite{Yuan2022QuantumMagnonics,Morris2017StrongCoupling}.

\section*{Acknowledgments} 

This work was supported by the U.S. National Science Foundation (NSF) under Awards No. ECCS-2426642 (W.Z. and B.Y.). Part of the work involving theoretical analysis was supported by the U.S. Department of Energy, Office of Science, Basic Energy Sciences under Award Number DE-SC0026305 (W.Z.). Y.L. is supported by the U.S. DOE, Office of Science, Basic Energy Sciences, Materials Science and Engineering Division under contract No. DE-SC0022060.

\section*{Data Availability} 

The data are available from the authors upon reasonable request.

\appendix

\section{Separation dependence of the inter-resonator coupling}
\label{app:distance_dependence}

\begin{figure}[htbp]
    \centering
    \includegraphics[width=\columnwidth]{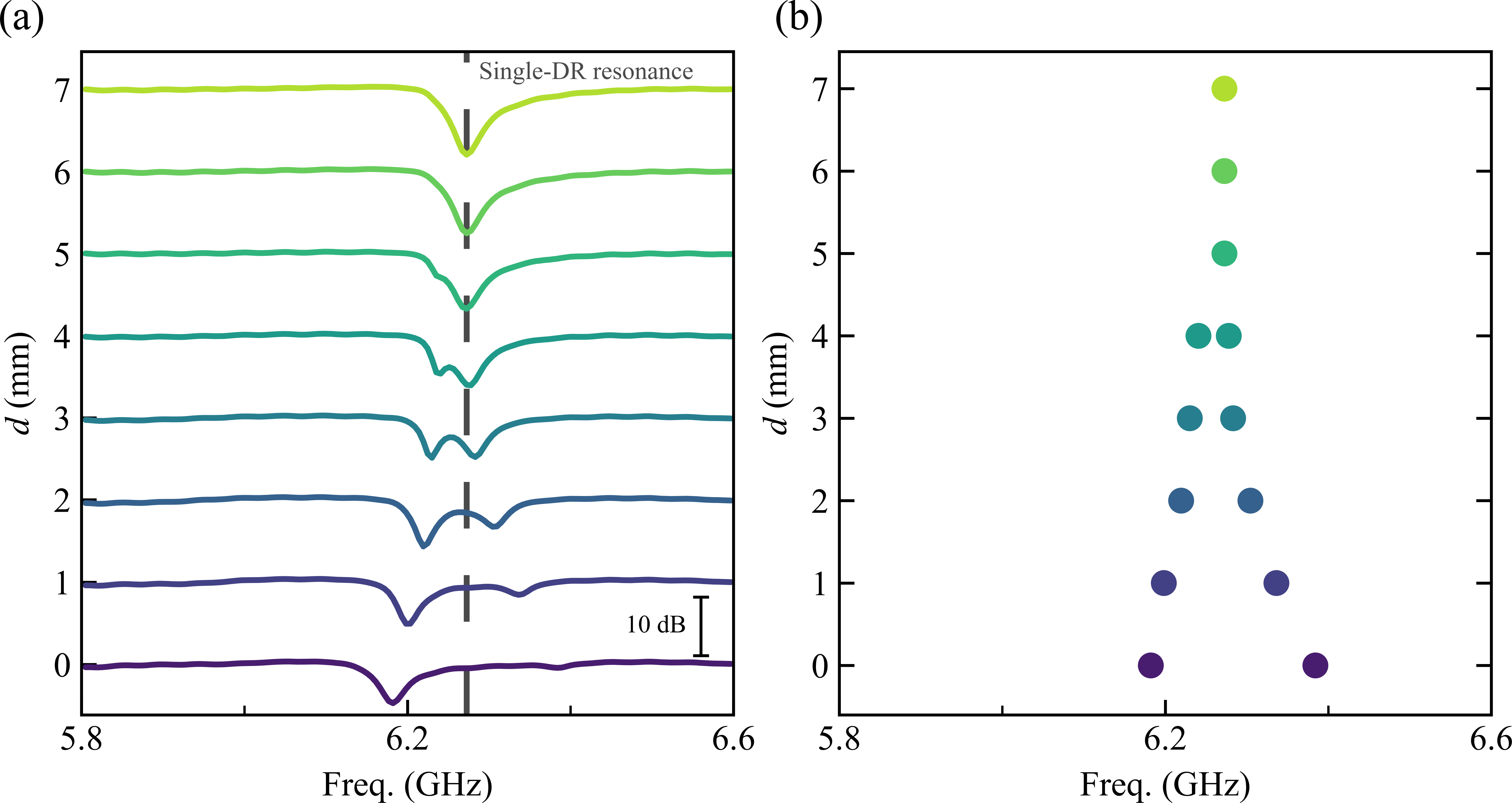}
    \caption{Separation dependence of the 2-DR transmission spectrum. (a) Zero-field transmission \(S_{12}\) in dB measured without the YIG sphere for edge-to-edge separations \(d=0\) to \(7~\mathrm{mm}\). The scale bar denotes a \(10~\mathrm{dB}\) offset to each transmission line. The vertical dashed line marks the single-DR resonance at \(6.273~\mathrm{GHz}\). (b) Resonance frequencies extracted from the spectra in (a).}
    \label{Appfig:distance_dependence}
\end{figure}

To experimentally establish the weak-coupling reference used in Fig.~\ref{fig:1}(b), we measure the transmission spectra of two identical DRs while varying their edge-to-edge separation \(d\) from \(0\) to \(7~\mathrm{mm}\). These measurements are performed without the YIG sphere and without an applied bias field. Fig.~\ref{Appfig:distance_dependence}(a) shows the resulting \(S_{12}\) spectra, vertically offset for clarity. The vertical dashed line marks the single-DR resonance at \(6.273~\mathrm{GHz}\).

At small separations, two distinct resonance minima are observed, indicating that the near-field interaction between the DRs lifts the degeneracy of their individual resonances and produces two collective photon modes. As \(d\) increases, the two resonances approach each other and their frequency separation progressively decreases. The corresponding resonance frequencies extracted from the spectra are summarized in Fig.~\ref{Appfig:distance_dependence}(b). Two resolvable resonance minima can be identified for \(d=0\)--\(4~\mathrm{mm}\). At \(d=5~\mathrm{mm}\), the lower-frequency feature is reduced to an asymmetric shoulder rather than a well-resolved second minimum and is therefore not included as an independent resonance in Fig.~\ref{Appfig:distance_dependence}(b). For \(d=6\) and \(7~\mathrm{mm}\), only a single resonance is resolved near the single-DR resonance frequency.

The absence of a resolvable doublet at \(d=6~\mathrm{mm}\), together with the similar single-resonance response at \(d=7~\mathrm{mm}\), indicates that the coupling-induced frequency separation at these distances is smaller than the experimental linewidth. We therefore use \(d=6~\mathrm{mm}\) as the weak-coupling reference in the main text, whereas the resolvable splitting at smaller separations establishes the formation of collective photon modes.

\section{Transmission spectrum and position-resolved coupling in the four-resonator array}
\label{app:4DR_map}

\begin{figure}[htbp]
    \centering
    \includegraphics[width=2.7 in]{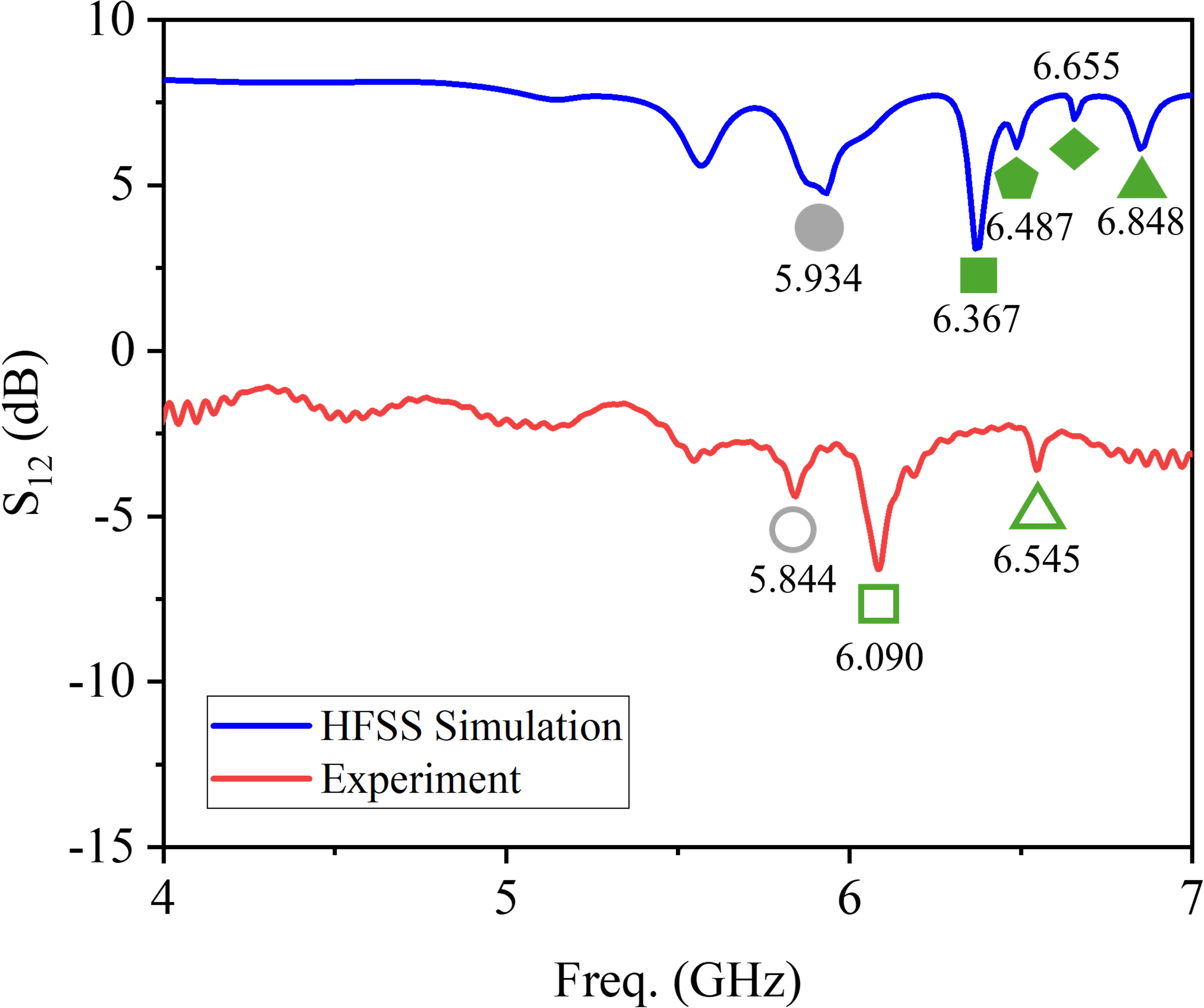}
    \caption{Measured zero-field transmission \(S_{12}\) of the compact 4-DR array (red), compared with the corresponding HFSS driven-response spectrum (blue). The measurement is performed without the YIG sphere and without an applied bias field, and the simulated spectrum is shifted vertically by \(+8.5~\mathrm{dB}\) for clarity. Solid and hollow symbols identify the simulated and experimental collective photon resonances, respectively, with matching shapes indicating the proposed mode correspondence.}
    \label{Appfig:4DR_S12}
\end{figure}

As a complementary test of the local-node picture and the limitations of the ideal uniform-chain model, we also examine the four-resonator array. We first identify its collective photon resonances by comparing the measured zero-field transmission \(S_{12}\) with the corresponding HFSS driven-response spectrum. The measurements are performed without the YIG sphere and without an applied bias field. As shown in Fig.~\ref{Appfig:4DR_S12}, the selected experimental resonances at \(6.090\), \(6.201\), \(6.373\), and \(6.545~\mathrm{GHz}\) are associated with the simulated modes at \(6.367\), \(6.487\), \(6.655\), and \(6.848~\mathrm{GHz}\), respectively, according to their frequency ordering. These resonances constitute the selected \(\mathrm{TE}_{01\delta}\)-derived collective-mode manifold used in the following position-resolved measurements.

\begin{figure*}[htp]
    \centering
    \includegraphics[width=6 in]{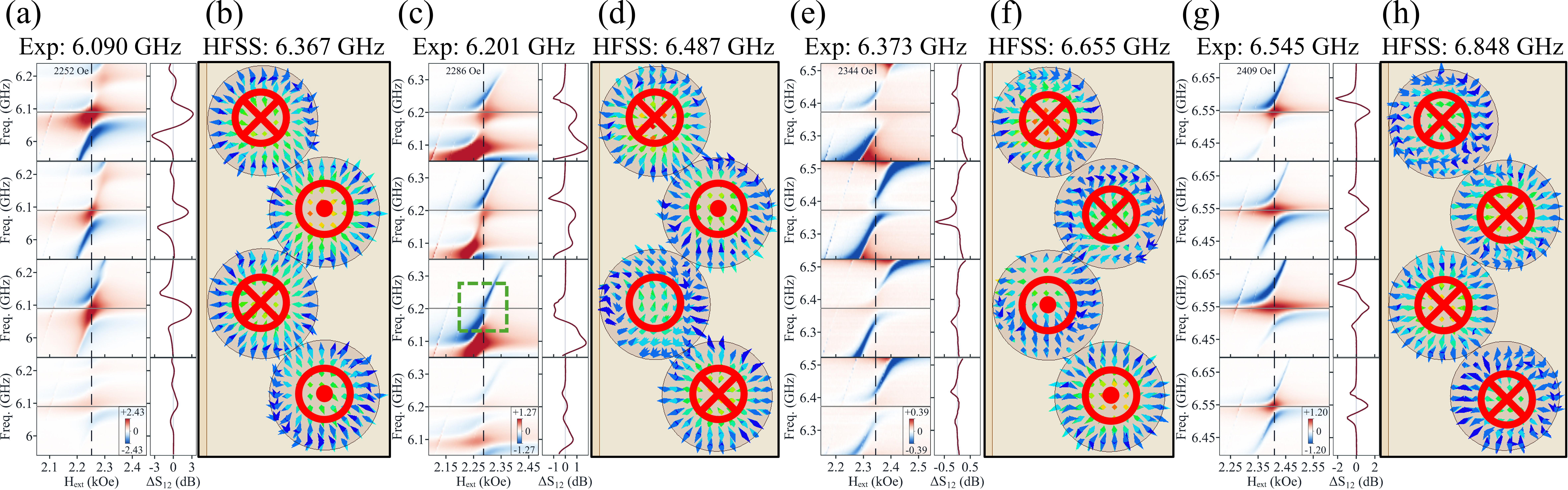}
    \caption{Position-resolved magnon--photon coupling in the four-resonator array. Differential transmission maps \(\Delta S_{12}\) measured near the four collective photon modes at (a) \(6.090\), (c) \(6.201\), (e) \(6.373\), and (g) \(6.545~\mathrm{GHz}\), together with the corresponding HFSS-calculated microwave magnetic-field profiles at (b) \(6.367\), (d) \(6.487\), (f) \(6.655\), and (h) \(6.848~\mathrm{GHz}\). Within each experimental group, the four rows from top to bottom correspond to the YIG sphere positioned directly above \(D_1\) through \(D_4\). The spectra extracted at the magnetic fields marked by the vertical dashed lines are shown beside the corresponding maps. The green dashed box in (c) identifies the \(6.201\)-GHz mode measured with the YIG sphere above \(D_3\), where no clear resolvable avoided crossing is observed. The HFSS field profiles are evaluated at a common RF phase. The \(\odot\), \(\otimes\), and \(\scalebox{0.8}{$\bigcirc$}\) symbols denote dominant local microwave magnetic fields directed along \(+z\), along \(-z\), and a strongly reduced \(z\)-directed component, respectively.}
    \label{Appfig:4_DR_map}
\end{figure*}
 
Having established the bare collective-mode frequencies and their experiment--simulation correspondence, we next compare the measured site dependence with the prediction of the ideal uniform-chain model. For an ideal uniform open chain of four resonators, the normalized eigenvector components are
\begin{equation}
    v_{n,j}^{(4)}
    =
    \sqrt{\frac{2}{5}}
    \sin\left[
    \frac{(5-n)j\pi}{5}
    \right],
    \qquad n,j=1,\ldots,4.
\label{eq:4DR_eigenvectors}
\end{equation}
None of the eigenvector components in Eq.~\ref{eq:4DR_eigenvectors} vanish at any of the four resonator sites. The ideal uniform-chain model therefore predicts no internal-node-induced suppression of the magnon--photon interaction in the four-resonator array. Provided that the local magnon--photon overlap is finite, every combination of collective mode \(n\) and YIG position \(D_j\) is expected to exhibit a nonzero effective coupling.

Fig.~\ref{Appfig:4_DR_map} shows the differential transmission maps measured near the four selected collective photon modes, together with the corresponding HFSS-calculated microwave magnetic-field profiles. Within each experimental frequency group, the four rows correspond to the YIG sphere positioned successively above \(D_1\), \(D_2\), \(D_3\), and \(D_4\). The spectra extracted at the magnetic fields indicated by the vertical dashed lines are displayed beside the corresponding maps.

Most of the measured mode--position combinations exhibit resolvable magnon--photon hybridization, consistent with the absence of exact internal nodes in the ideal four-resonator eigenvectors. An exception occurs for the \(6.201\)-GHz mode when the YIG sphere is positioned above \(D_3\), as marked by the green dashed box in Fig.~\ref{Appfig:4_DR_map}(c). At this position, no clear resolvable avoided crossing is observed, even though the ideal eigenvector has a nonzero component \(v_{2,3}^{(4)}\). Moreover, the ideal model gives \(\lvert v_{2,2}^{(4)}\rvert=\lvert v_{2,3}^{(4)}\rvert\), whereas the measured coupling responses above \(D_2\) and \(D_3\) are markedly different. The experimental suppression at \(D_3\) is therefore not captured by the scalar eigenvector of the ideal uniform chain.

The corresponding HFSS profile provides additional information about this deviation. For the \(6.487\)-GHz simulated mode [Fig.~\ref{Appfig:4_DR_map}(d)], the local microwave magnetic field at \(D_3\) is predominantly in plane, with only a limited projection along \(z\), consistent with the absence of a clear avoided crossing in the corresponding \(6.201\)-GHz measurement. Thus, the experiment and the full-wave simulation exhibit the same site-selective suppression, while both deviate from the prediction of the ideal uniform-chain model. The suppressed interaction at \(D_3\) should consequently be interpreted as a reduction of the relevant local field component and magnon--photon overlap, rather than as an exact node of the total microwave magnetic field.

A likely origin of this deviation is the geometrical asymmetry of the four-resonator arrangement relative to the microstrip line. In the experimental and simulated configurations, the resonators on opposite sides of the array are not positioned equivalently with respect to the microstrip. This asymmetry breaks the mirror symmetry assumed in the ideal model and modifies the local field polarization, microwave loading, dissipation, and magnon--photon overlap. Consequently, equal eigenvector amplitudes in the scalar uniform-chain model do not necessarily produce equal \(z\)-directed microwave magnetic fields or equally resolvable avoided crossings at nominally symmetry-related sites. A quantitative description of this effect will require an extended model incorporating the vector character of the microwave field together with mode- and site-dependent coupling and loss parameters.

\section{Generalized theory of \(N\)--resonator arrays}
\label{app:N_DR_generalization}

For completeness, the local coupling selection rule derived for the three-resonator and five-resonator arrays can be generalized to an open chain containing an arbitrary number \(N\) of nominally identical resonators. Under the same nearest-neighbor approximation used in the main text, the photon Hamiltonian is
\begin{equation}
    \frac{\hat H_{\mathrm{ph}}^{(N)}}{\hbar}
    =
    \omega_0\sum_{j=1}^{N}\hat c_j^\dagger\hat c_j
    +
    J\sum_{j=1}^{N-1}
    \left(
    \hat c_j^\dagger\hat c_{j+1}
    +
    \hat c_{j+1}^\dagger\hat c_j
    \right).
\label{eq:N_dr_photon_hamiltonian}
\end{equation}

Diagonalization of this uniform open chain gives \(N\) collective photon operators,
\(\hat a_n=\sum_jv_{n,j}^{(N)}\hat c_j\). Using the same sign convention as in the three-resonator model and indexing the modes in ascending frequency for \(J>0\), the eigenfrequencies and normalized eigenvector components are
\begin{equation}
\begin{aligned}
    \Omega_n^{(N)}
    &=
    \omega_0-2J\cos\left(\frac{n\pi}{N+1}\right),\\
    v_{n,j}^{(N)}
    &=
    \sqrt{\frac{2}{N+1}}
    \sin\left[
    \frac{(N+1-n)j\pi}{N+1}
    \right],\\
    \qquad n,j&=1,\ldots,N.
\end{aligned}
\label{eq:N_dr_eigenmodes}
\end{equation}

When the YIG sphere is positioned above resonator \(D_j\), projecting the local magnon--photon interaction onto collective photon mode \(n\) gives
\begin{equation}
    G_{n,D_j}^{(N)}
    =
    g_jv_{n,j}^{(N)},
    \qquad
    v_{n,j}^{(N)}=0
    \ \Longrightarrow\
    G_{n,D_j}^{(N)}=0.
\label{eq:N_dr_selection_rule}
\end{equation}
Here, \(g_j\) contains the local microwave-field overlap and polarization matching between the Kittel mode and the photon mode associated with \(D_j\). An exact node of the collective eigenvector therefore produces a vanishing local coupling in the ideal model. Conversely, a nonzero eigenvector component permits magnon--photon hybridization but does not require equal coupling strengths at different sites, because \(g_j\) may vary with the resonator position and microwave-field environment. Equations~\ref{eq:N_dr_eigenmodes} and \ref{eq:N_dr_selection_rule} contain the three- and five-resonator results discussed in the main text as specific cases and provide the node-based coupling prediction for an arbitrary uniform resonator chain.

\section*{References}

\bibliography{ref}

\end{document}